# GRAD: On Graph Database Modeling

Amine Ghrab
*EURA NOVA R&D, Belgium*
amine.ghrab@euranova.eu

Oscar Romero
*Universitat Politècnica de Catalunya, Spain*
oromero@essi.upc.edu

Sabri Skhiri
*EURA NOVA R&D, Belgium*
sabri.skhiri@euranova.eu

Alejandro Vaisman
*Instituto Tecnológico de Buenos Aires, Argentina*
avaisman@itba.edu.ar

Esteban Zimányi
*Université Libre de Bruxelles, Belgium*
ezimanyi@ulb.ac.be

*Abstract.* *Graph databases have emerged as the fundamental technology underpinning trendy application domains where traditional databases are not well-equipped to handle complex graph data. However, current graph databases support basic graph structures and integrity constraints with no standard algebra. In this paper, we introduce GRAD, a native and generic graph database model. GRAD goes beyond traditional graph database models, which support simple graph structures and constraints. Instead, GRAD presents a complete graph database model supporting advanced graph structures, a set of well-defined constraints over these structures and a powerful graph analysis-oriented algebra.*

## 1. INTRODUCTION

Graphs have been an active field of study for decades (Newman, 2010). A large repertoire of algorithms and frameworks had been developed for efficient graph analysis. The great expressive power of graphs encouraged their use for modeling domains presenting complex structural relationships. In biology for example, graphs are used for modeling metabolic pathways, genetic regulations and protein interactions (Zimányi & Skhiri, 2004). In social networks, they are used for modeling relationships between users and for mining structural information such as churn prediction (Dasgupta et al., 2008) or marketing strategies (Gomez Rodriguez et al., 2010). On the other hand, a multitude of emerging problems are naturally represented using graph models and solved using graph algorithms. Business problems such as fraud detection, trend prediction, product recommendation, network routing and optimization just to name a few, are solved using graph algorithms (Skhiri & Jouili, 2013; Han et al., 2011). For example, by examining the betweenness centrality of a node in a social network (i.e., the number of shortest paths between all pairs of nodes that traverse a given node), an analyst can detect influential information brokers. For a telecommunication operator, the betweenness centrality is used to detect central routers or antennas, and thus contribute in optimizing the routing and load balancing across the infrastructure.

The wide range of important domains and applications on graphs calls for the development of novel techniques and industry-grade systems for efficient management of graph data. A plethora of frameworks were introduced for the processing, management and analysis of graph data (Angles & Gutierrez, 2008; Martinez-Bazan et al., 2007; Shao et al., 2013). Among these frameworks, graph databases are emerging as the fundamental component for the storage and querying of graph data. Graph databases are built based on a database model, which provides the required solid theoretical foundations. A database model, as defined by Codd (1980) consists of a set of (1) data structures, (2) integrity constraints, and (3) manipulation operators.

Graph data management systems are commonly implemented using specialized and native graph databases engines and models (Shao et al., 2013). Native graph databases have the advantage of avoiding or at least reducing the impedance mismatch problem (Sadalage & Fowler, 2012). In the context of data management systems, impedance mismatch appears at three levels: modeling, querying and programming levels. Therefore, as a first benefit of using a native graph database, the modeling phase flows naturally, since the structures are similar to the way end-users perceive the domain. Querying is also user-friendly and less error-prone than in the first approach, since the operators target directly the network structure and can examine specific graph measures. Moreover, this approach avoids transformation and mapping problems at both design and querying phases. For instance, operations such as traversals are easier to express and are executed at a much lower cost in native graph databases than through joins in relational databases. Finally, while advanced concepts such as assertions are not implemented on current relational databases, they can be defined and implemented in graph databases by using graph patterns as we will show later in this paper.

Real-world graph data are often dynamic, irregular (i.e., they do not conform to a rigid schema), and heterogeneous (multiple types of nodes and edges with arbitrary properties exist within the same graph). For instance, some graph elements describing similar real-world concepts might have different properties (i.e., missing or extra attributes), and new attributes or types of relationships could be added on-the-fly. Rigid schemas are therefore not well-suited for capturing and efficiently analyzing graph data. A loose and non-strict schema provides better support to dynamic graph data, where the changes in both content and structure are frequent. It is well-known that schema update and maintenance of the schema are complex and time-consuming operations in relational databases. On the other hand, most of the current graph databases adopt a weak notion of schema or do not handle it at all (i.e., they assume the schema to be application-implicit). Usually, the boundary between the schema and instance data is blurred and no strict data types or schemas are cast on the graph.

In this paper we address the need for a native and complete (according to Codd (1980)), graph database model, introducing GRAD (GRAph Database model). GRAD is a novel and generic database model designed for advanced modeling and rich analysis of graphs. It is equipped with special graph data structures, shown in Figure 1 (e.g., hypernodes, see the GRAD data model section), that natively support common analytical operations such as data integration and encapsulation. GRAD is a schema-less data model. Graph data remain self-descriptive, and analysis is mainly done through traversal and neighborhood exploration. In GRAD, nodes and edges are first-class citizens, and the model is equipped with content and topology-aware operators that handle entire graphs. The model fits the "*load first, model later*" data management strategy (Olsson, 2013), which fosters an on-read schema, which is more convenient for irregular dynamic graph data than the traditional on-write schema. The idea is to load the data into a graph repository without restricting it to any predefined schema. At analysis time, the analyst could design an application-implicit, on-read schema and then have the flexibility to explore more conveniently the graph at run-time according to the specific analysis scenario. Moreover, GRAD strengthens integrity of the graph database through constraints defined over the graph structures and maintained through the different algebraic operations. Finally, GRAD can be implemented on top of various underlying database engines.

In summary, our contributions are:

- A novel graph database model that uses advanced data structures that capture the common notions present in typical conceptual data modeling languages.

- A set of rules to enforce and preserve the integrity of the graph data, with a focus on graph entity integrity and semantic constraints.

- A collection of algebraic operators to enable online graph querying and analysis. Graphs are the operands and the return type of all the operators.

- An implementation of GRAD on top of a Neo4j graph database, and an architecture which shows that GRAD can be implemented on a wide range of data management systems.

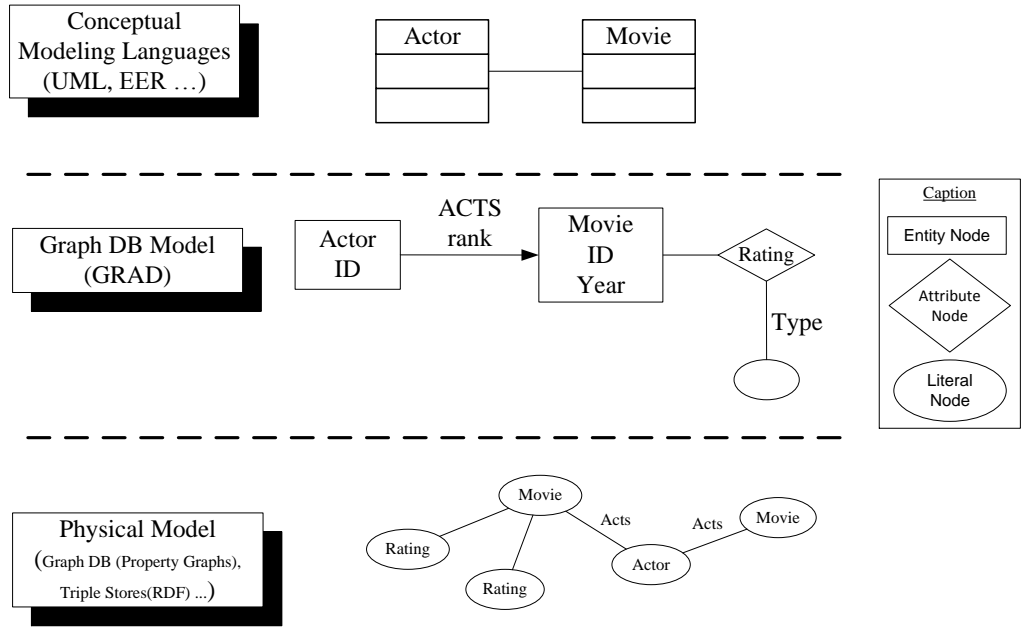

Figure 1. Positioning of GRAD on the data modeling landscape

The remainder of this paper is organized as follows. In the next section, we introduce the motivation and running example used along the paper. Then, we present the advanced data structures of GRAD where we project the common notions form conceptual modeling languages into the graph setting. We then define the different integrity constraints ensuring the graph database integrity. Following this, we present the GRAD algebra for querying and manipulating the graph data. Applications of GRAD on graph management and the architecture of GRAD-based graph analysis framework are presented afterwards. Finally, related work is discussed and we conclude the paper and describe future work.

## 2. RUNNING EXAMPLE

We next introduce the running example that we use throughout the paper. The data used in this paper come from the MovieLens + IMDb/Rotten Tomatoes dataset, which was first published in the HetRec2011 workshop (Cantador et al., 2011). In the sequel we refer to this dataset simply as *MovieLens*. The dataset integrates information about movies from MovieLens[1] with their corresponding web pages at Internet Movie Database (IMDb)[2] and Rotten Tomatoes[3] movie review systems. We represent this dataset using property graphs as illustrated in Figure 2. The network contains information about movies classified by genres, and filmed across multiple locations. Movies have different attributes, such as year of release, titles, ratings and scores from different communities, and are linked to directors and actors (ranked by importance).

---

[1] http://grouplens.org/datasets/movielens/
[2] http://www.imdb.com
[3] http://www.rottentomatoes.com

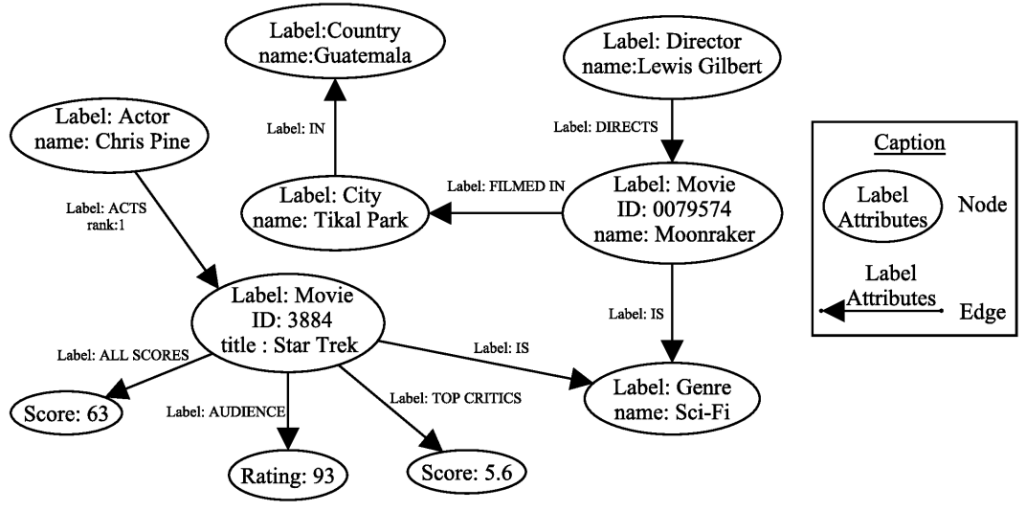

Figure 2. MovieLens represented with Property Graphs

## 3. PRELIMINARIES

Graph data are usually represented as collections of nodes and edges. Nodes represent entities and edges relate pairs of nodes, and are used to represent different types of relationships. Nodes and edges might be labeled, and have a set of properties represented as attributes. Many current graph databases represent graphs using the *property graph* model. Property graphs describe a directed, labeled and attributed multi-graph (Rodriguez & Neubauer, 2010). We formally define a property graph as follows:

**Definition 1** (Property Graph) A ***property graph*** is a tuple $\mathcal{G} = (\mathcal{V}, \mathcal{E}, \mathcal{L}_v, \mathcal{F}_v, \mathcal{L}_e, \mathcal{F}_e, \Lambda_v, \mathcal{H}_v, \Lambda_e, \mathcal{H}_e)$, where:

- $\mathcal{V}$ is a set of nodes.
- $\mathcal{E} \subseteq \mathcal{V} \times \mathcal{V}$ is a set of edges.
- $\mathcal{L}_v$ is a set of node labels and $\mathcal{F}_v: \mathcal{V} \to \mathcal{L}_v$ is the function that assigns a label to each node.
- $\mathcal{L}_e$ is the set of labels on entity edges and $\mathcal{F}_e: \mathcal{E} \to \mathcal{L}_e$ is the function that assigns a label to each edge.
- $\Lambda_v = \{a^1, a^2, \ldots, a^n\}$ is a set of node attributes represented as key/value pairs. Each node is associated with an attribute vector $[a^1, a^2, \ldots, a^k]$.
- $\mathcal{H}_v = \{h_v^1, h_v^2, \ldots, h_v^n\}$ is a set of $n$ functions such that: $h_v^i: \mathcal{V} \to dom(a^i)$. For a node $v_j \in \mathcal{V}$, the function $h_v^i(v_j)$ returns the value $x$ of its $i$-th attribute, and $x \in dom(a^i)$.
- $\Lambda_e = \{b^1, b^2, \ldots, b^m\}$ is a set of edge attributes represented as key/value pairs. Each edge is associated with an attribute vector $[b^1, b^2, \ldots, b^k]$.
- $\mathcal{H}_e = \{h_e^1, h_e^2, \ldots, h_e^m\}$ is a set of functions such that: $h_e^i: \mathcal{E} \to dom(b^i)$. For an edge $e_j \in \mathcal{E}$, the function $h_e^i(e_j)$ returns the value $x$ of its $i$-th attribute, and $x \in dom(b^i)$.

Current leading graph databases (e.g., Neo4j[4] and Titan[5]) are built on top of property graphs. However, property graphs do not define a full database model but rather describe basic data structures which are simple and oriented for storage and operational workloads. A native and comprehensive graph database model oriented to graph analysis and compliant to Codd's definition (Codd, 1980) needs to be designed. For advanced graph modeling and analysis, semantically richer graph structures are needed, as well as the integrity constraints and manipulation operators. We address these needs in the next section.

## 4. THE *GRAD* DATA MODEL

Multiple conceptual data models exist in the literature to support and standardize the representation of different domains and applications. For each of these conceptual models, a language is defined. Currently, the most used languages are the Unified Modeling Language (UML), and the Extended Entity-Relationship Model (EER). A recent study by Keet et al. (Keet & Fillottrani, 2013) shows that these languages have some common core entities. For example, *object type, relationship and attribute* are concepts present in most current modeling languages, although naming conventions might differ. We map these concepts to GRAD and formally define them using the graph data structures as depicted in Figure 1. This mapping shows GRAD's ability to capture traditional modeling concepts and to project them on graphs. From a designer's perspective, the mapping simplifies the task of representing familiar concept from modeling languages using graphs as first-class citizens.

The GRAD graph model extends property graphs by assigning specific semantics to graph nodes and edges. This is achieved by means of complex attributes on the nodes, and explicit annotations on edges. We next present formally the GRAD graph model.

**Definition 2** (GRAD Graph) A *GRAD graph* is a tuple $\mathcal{G} = (\mathcal{V}, \mathcal{E}, \mathcal{L}_v, \mathcal{F}_v, \mathcal{L}_e, \mathcal{F}_e, \Lambda_v, \mathcal{H}_v, h_l, \Lambda_e, \mathcal{H}_e)$, where:

- $\mathcal{V} = (V_e \cup V_a \cup V_l)$ is a set of nodes, where $V_e$ is a set of entity nodes, $V_a$ a set of attribute nodes, and $V_l$ a set of literal nodes.

- $\mathcal{E} = (E_e \cup E_a \cup E_l)$ is a set of edges, where $E_e$ is a set of entity edges, $E_a$ a set of attribute edges, and $E_l$ a set of literal edges.

- $\mathcal{L}_v$ is a set of labels on entity and attribute nodes. $\mathcal{F}_v: (V_e \cup V_a) \to \mathcal{L}_v$ is a function that assigns a label to each entity or attribute node.

- $\mathcal{L}_e$ is a set of labels on entity edges and $\mathcal{F}_e: E_e \to \mathcal{L}_e$ is a function that assigns a label to each entity edge.

- $\Lambda_v = \{a^1, a^2, \ldots, a^n\}$ is a set of entity node identifiers. Each node is associated with a vector of identifiers: $[a^1, a^2, \ldots, a^k]$.

- $\mathcal{H}_v = \{h_v^1, h_v^2, \ldots, h_v^n\}$ is a set of *n* functions such that: $h_v^i: V_e \to dom(a^i)$. For a node $v_j \in V_e$, the function $h_v^i(v_j)$ returns the value *x* of its *i-th* attribute, where $x \in dom(a^i)$.

- $h_l: V_a \times E_l \to dom(val)$ is the function that returns the value *val* stored for literal nodes. The function $h_l$ returns the actual value *val* of an attribute given the attribute node and the literal edge associated to the literal node storing *val*.

- $\Lambda_e = \{b^1, b^2, \ldots, b^m\}$ is the set of edge attributes represented as key/value pairs. Each edge is associated with an attribute vector $[b^1, b^2, \ldots, b^k]$.

---
[4] http://neo4j.org
[5] http://thinkaurelius.github.com/titan/

- $\mathcal{H}_e = \{h_e^1, h_e^2, \ldots, h_e^m\}$ is the set of m functions such that: $h_e^i: E_e \cup E_l \to dom(b^i)$. For an edge $e_j \in (E_e \cup E_l)$, the function $h_e^i(e_j)$ returns the value $x$ of its $i$-th attribute, where $x \in dom(b^i)$.

The first key concept present in modeling languages that we need to define on GRAD, is the concept of *class*. In modeling languages, a class is used to represent a set of things of a specific kind that share a common structure and relationships. A class is characterized by a predicate, such that all elements belonging to the same class satisfy the class's predicate. For the sake of simplicity, we limit the discussion to unary predicates, although more complex predicates could be defined and applied as well. Formally, we define a class of graph elements as follows:

**Definition 3** (*Class*) A *class* $\Sigma_i$ in GRAD describes a set of graph elements that satisfy a unary predicate applied on the labels of entity nodes. Each class $\Sigma_i$ is characterized by a label $C_i$. Therefore, $\forall v \in V_e$, where v belongs to the class $\Sigma_i$, characterized by the label $C_i$, iff $\mathcal{F}_v(v) = C_i$.

The set of k classes on a graph is disjoint (i.e., $\forall i, j, \Sigma_i \cap \Sigma_j = \emptyset$), and is denoted as $\Sigma = \{\Sigma_1, \Sigma_2, \ldots, \Sigma_n\}$. $C = \{C_1, C_2, \ldots, C_n\}$ is the set of all labels characterizing classes on $\Sigma$, with $C \subseteq \mathcal{L}_v$.

**Example 1** Classes in *MovieLens* could be *MOVIE, ACTOR, USER* etc. The predicate of the class describing movies is expressed as: $\mathcal{F}_v(v)=MOVIE$.

The second key concept from modeling languages that we must represent using GRAD is the notion of *object*. Objects describe concrete concepts (i.e., real-world entities) by means of unique identifiers, a set of attributes describing their state, and a set of relationships. Each object belongs to only one class at a time, and objects satisfying the same predicate are grouped in the same class. In GRAD, the core of an object is represented by the entity node which contains the label and the identifier attributes of the real-world entity. Attribute nodes are attached to entity nodes and denote non-identifier attributes of the object, and might be multi-valued. Literal nodes record the actual value of their corresponding attribute node each time a new value is added. We formally define each of these concepts as follows:

**Definition 4** (*Entity Node*) An *entity node* $v_i \in V_e$ is a pair $<C_i, ID_i>$ where $\mathcal{F}_v(v_i) = C_i$, such that $C_i \in \mathcal{L}_v$ denotes the "type" of the entity node (i.e., the class to which it belongs), and $ID_i$ is the set of identifiers that unequivocally identify the entity node among the nodes of the same class. Each element of the set $ID_i$ is immutable and might be simple or composite.

Each entity node represents the core part of the information about the real-world graph element. The concept of *node identifier* maps to the concept of *primary key* in databases and is considered to be domain-related and not system-generated.

**Example 2** An entity node of the class *MOVIE* could be represented with the pair *<MOVIE, {3884, Star Trek}>*. Here, the movie identifier is composed of the identifiers on the source websites IMDB and Rotten Tomatoes.

Attributes of real-world entities might have different values according to context change or new data integration procedures (e.g., updates, time-stamping etc.). We capture these changes using attribute and literal nodes, defined next.

**Definition 5** (Attribute Node) An *attribute node* contains only a label describing its name. An attribute node $v_j \in V_a$ should be associated to an entity node v and is defined by its label $l_j$. The set of labels of the k attribute nodes of v is denoted as $L_a = \{l_1, l_2, \ldots, l_k\}$. Note that an attribute node denotes an attribute of the entity node that is not an identifier and may have multiple values.

The set of attribute nodes on the graph is denoted $V_a$ and the set of labels of all attribute nodes is denoted $\mathcal{L}_a$, where $\mathcal{L}_a \subseteq \mathcal{L}_v$. Note that $\mathcal{L}_v = C \cup \mathcal{L}_a$. An attribute node must be linked to exactly one entity node.

**Example 3** In the *MovieLens* network, an attribute node associated with a movie entity node may be its score, the title of a movie provided in multiple languages, or its identifier on each website. In that case, the movie entity node will have a set of associated attribute nodes $L_a(MOVIE)$ = {*Score, Title, WebID*}.

The actual value of an attribute is represented using literal nodes. GRAD captures the concept of multivalued attribute (as defined on typical EER models) using an attribute node with all its related literal nodes.

**Definition 6** (Literal Node) A *literal node* $v \in V_l$ captures the value (denoted as *val*) of its corresponding attribute node ($u \in V_a$) in a given context. This value could be simple or composite. The function $h_l: V_a \times E_l \rightarrow dom(val)$ is used to return the actual value $val = h_l(u, e)$ stored in v.

**Example 4** An attribute node labeled "*title*" is attached to movie nodes to represent the movie title in different languages. The actual new title is added as a literal node (e.g., the title of the movie translated from English to French).

We next study the types of relationships that could link the pairs of nodes of the kinds defined above. In conceptual modeling languages (e.g., UML) various kinds of relationships between classifiers are supported, such as (1) Association, (2) Generalization and (3) Dependency. Aggregation (representing part-whole relationship) and Composition (representing ownership relationships, with lifetime dependency) are treated as specific associations, while Realization and Usage are sub-types of Dependency (Rumbaugh et al., 2004). Similarly to YAM[2] (Abelló et al., 2006), GRAD is focused on data modeling, therefore we do not consider dependency relationships since they are mainly designed for application modeling purposes. In addition, in this paper we consider only binary relationships between pairs of nodes, although hypergraphs could be used to model edges connecting an arbitrary number of nodes (Iordanov, 2010). Regarding the nodes they link, we classify edges as *entity*, *attribute* and *literal* ones, defined next.

**Definition 7** (Entity Edge) An *entity edge* $e_i \in E_e$ between a pair of entity nodes $v_m, v_n \in V_e$ is defined as: $e_i = (v_m, v_n, type_i, l_i, B_i)$, where $v_m$ is called the start-node and $v_n$ is the end-node. $l_i \in \mathcal{L}_e$ is the entity edge's label and $B_i$ = {$b_1, b_2, ...b_n$} is the set of its attributes; $type_i$ denotes the specific type of the relationship, with $type_i \in$ {Association, Generalization, Aggregation, Composition}.

To allow richer semantics to be expressed on the graph, we define four types of entity edges as follows:

- Association edge: This is the most common type of relationships, where two entity nodes are associated with each other. Association edges have the same meaning of the association concept in UML. The set of association edges is denoted $E_{as}$.

- Generalization edge: This relationship relates a subclass entity node to its superclass entity node. It is usually referred to as an "Is A" relationship, and maps to generalization in UML. Each entity node can be generalized to at most one superclass (i.e., have at most one outgoing generalization edge from the subclass to the superclass). The set of composition edges is denoted $E_g$.

- Aggregation edge: Describes a whole/part relationship between two entity nodes. In UML this maps to the concept of (shared) aggregation and reflects the weak dependency between the two entities. Each entity node could be part of at most one whole element (i.e., have at most one outgoing aggregation edge from the part to the whole). This type of edges describes a hierarchical relationship between data entities, such as the relationship between cities and countries. The set of aggregation edges is denoted as $E_{ag}$.

- Composition edge: This is a stronger form of aggregation that reflects that the existence of the part depends on the existence of the composite (whole) entity node. Therefore, if the node representing the whole is deleted, the part is consequently deleted. We describe the entity node representing the part as a weak entity node that can only exist if its composite exists. Each entity node could be included as a part of at most one composite (i.e., have at most one outgoing composition edge from the part to the whole). This relationship maps to the composition (composite aggregation) in UML. The set of composition edges is denoted as $E_c$.

An entity edge must be typed using one of the four types defined above: $E_e = E_{as} \cup E_g \cup E_{ag} \cup E_c$. Moreover, each entity node might have many association edges, but at most one outgoing generalization (similarly aggregation or composition) edge. We use the many-to-one property of these relationships (i.e., generalization, aggregation and composition) to introduce the notion of parent-child relationship from the part to the whole, and from the subclass to the superclass.

**Example 5** As an example of association edges, relationship between movies and actors is represented by an entity edge labeled "*ACTS*" and having the attribute *ranking* to denote the actor's ranking in the movie. For generalization edges, if we imagine adding the class "*crew of a movie*" to the network, then both actors and directors belong to the "*Crew*" class. Finally, the relationship between cities and countries is a composition, because whenever a country is removed, all of its cities are removed as well.

In addition to entity edges, we introduce two other types of edges where the end nodes could be attribute and literal nodes. Attribute edges keep track of the changing attributes extracted as new nodes and are defined next.

**Definition 8** (Attribute Edge) An ***attribute edge*** represents a composition relationship (i.e., a life-cycle dependency where the same properties introduced earlier for composition relationships are valid for this edge also) between an attribute node and its parent entity node. Attribute edges do not embed attributes or labels. The set of attribute edges is denoted $E_a \subseteq V_e \times V_a$.

**Definition 9** (Literal Edge) A ***literal edge*** reflects a composition relationship between a literal node (part) and its parent attribute node (composite). Each literal edges has a set of attributes and $B_l = \{b_1, b_2, ...b_m\}$. Each attribute describes a part of the context for the value on the literal node. The set of literal edges is denoted as ($E_l \subseteq V_a \times V_l$).

**Example 6** A literal edge might indicate the title of the movie in a given language, while the title itself is stored in the related literal node.

Following our definition of composition, we describe attribute (resp. literal) nodes as weak nodes (i.e., representing weak entities) dependent on the entity (resp. attribute) node (representing the strong entity) to which they are linked by their attribute (resp. literal) edge. We also preserve the general notation to minimally denote an edge *e* directed from a node $v_a$ to a node $v_b$ as: $e = (v_a, v_b)$.

The graphical representation of the different types of nodes and edges as GRAD's data structures is depicted in Figure 3. GRAD graphs are exclusively represented by the graph structures introduced in this section, thus preserving the attributes and the topological properties of each element.

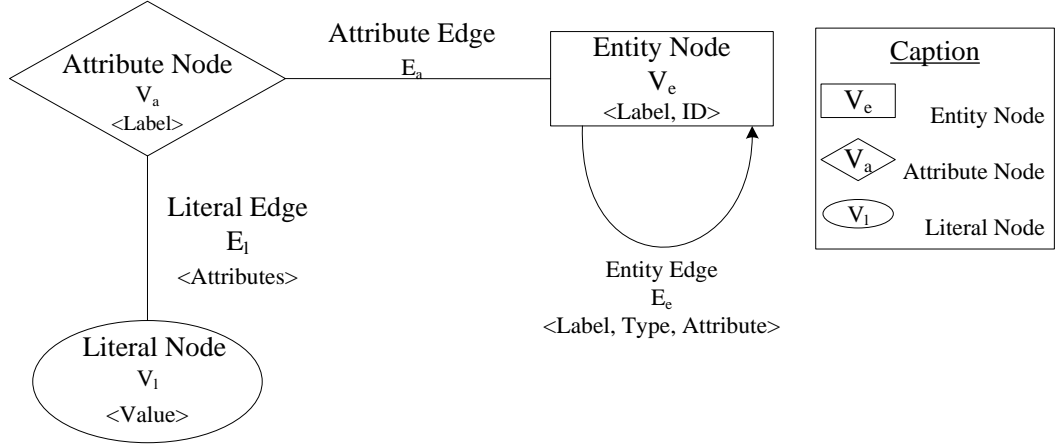

Figure 3. Graphical notation of GRAD structures

We use the concept of hypernode to fully represent a real-world object. Hypernodes are useful for representing various concepts such as aggregation or encapsulation of graph elements, providing a coarser view of the graph as it enables a higher level design and analysis of the network. We represent hypernodes on GRAD using subgraphs. In graph theory, G' is called a subgraph of G if G' contains a subset of nodes (V') and a subset of edges (E') of G. If G' preserves all the edges, originally present in G between the subset of nodes V', then the subgraph is called induced, otherwise it is called a partial subgraph. Therefore, a hypernode in GRAD is an induced subgraph of the data graph. It is represented as a two-level tree where the entity node is the root and the literal nodes are the leaves. Each hypernode groups an entity node, all its attribute and literal nodes, and the edges between them. With the graph data structures formally defined above, we define a GRAD hypernode as follows:

**Definition 10** (GRAD Hypernode) A ***hypernode*** in GRAD (whose entity node is $v_i \in V_e$) is defined formally as a subgraph $\Gamma_{v_i} = (V, E)$, where $V = \{v_i \cup V_{a_i} \cup V_{l_i}\}$ and $E = \{E_{a_i} \cup E_{l_i}\}$. $V_{a_i} \in V_a$ is the set of all the attribute nodes attached to $v_i$, and $V_{l_i} \in V_l$ is the set of all the literal nodes attached to attribute nodes of $V_{a_i}$. Similarly, $E_{a_i} \in E_a$ (resp. $E_{l_i} \in E_l$) are the edges linking the attribute nodes from $V_{a_i}$ to $v_i$ (resp. linking nodes form $V_{l_i}$ to their attribute nodes from $V_{a_i}$). The label of an entity node defines the class to which its encapsulating hypernode belongs.

**Example 7** Each movie in *MovieLens* is represented by a hypernode that contains the entity node representing the movie and its ranking as an attribute node. The ranking keeps track of the different values of the movie rating by community.

Note that each node, attribute edge and literal edge, are part of only one hypernode: $\forall u \in V$ (resp., $e \in E_a \cup E_l$), $\exists! \Gamma_v \mid u \in \Gamma_v$ (resp., $e \in \Gamma_v$). All data on the graph belonging to a given real-world entity are gathered and integrated in one hypernode, which simplifies the data integration tasks to be performed later.

**Example 8** A subgraph of the *MovieLens* network G is depicted in Figure 4 and is represented using GRAD structures as follows:

- $C$ = {MOVIE, CITY, COUNTRY, ACOTR, DIRECTOR}, $\mathcal{L}_a$ = {Rating}.

- $V_e = \{v_1, v_2, v_3, v_4, v_5, v_6\}$, $V_a = \{v_{Rating}\}$ and $V_l = \{v_{aud}\}$.

- $\mathcal{F}_v(v_1)$ = "MOVIE", $\mathcal{F}_v(v_2)$ = "CITY", $\mathcal{F}_v(v_3)$ = "COUNTRY", $\mathcal{F}_v(v_4)$= "DIRECTOR", $\mathcal{F}_v(v_5)=\mathcal{F}_v(v_6)$ = "ACTOR", $\mathcal{F}_v(v_{Rating})$ = "Rating".

- $E_{as} = \{e_{12}, e_{14}, e_{15}, e_{16}\}$, $E_c = \{e_{23}\}$, $E_a = \{e_{1r}\}$, and $E_l = \{e_{rv}\}$.

- $\mathcal{L}_e$ = {ACTS, DIRECTS, FILMED IN, LOCATED IN}.
- $\mathcal{F}_e(e_{12})$ = "FILMED IN", $\mathcal{F}_e(e_{23})$="LOCATED IN", $\mathcal{F}_e(e_{14})$= "DIRECTS", and $\mathcal{F}_e(e_{15})=\mathcal{F}_e(e_{16})$="ACTS".
- $\Lambda_v$ = {IMDB_ID, RT_ID, CityName, CountryName, DirectorName, ActorName}.
- $h_v^1(v_1)$ = "3884", $h_v^2(v_1)$ = "Star_Trek", $h_v^3(v_2)$ = "UTAH", $h_v^4(v_3)$ = "USA", $h_v^5(v_4)$="J.J._Abrams", $h_v^6(v_5)$ = "Eric_Bana", $h_v^6(v_6)$ $(v_6)$ = "Chris_Pine".
- $h_i(v_{Rating}, e_{rv})$ = "8.5".
- $\Lambda_e$ = {ranking, Type} and $h_e^1(e_{14})$ = "1", $h_e^1(e_{15})$ = "1", $h_e^2(e_{rv})$ = "Audience".
- $v_1$ = (MOVIE, {3884, Star_Trek}), $v_2$ = (CITY, {UTAH}), $v_3$ = ( COUNTRY, {USA}), $v_4$ = (DIRECTOR, {J.J._Abrams}), $v_5$ = (ACTOR, {Eric_Bana}), $v_6$ = (ACTOR, {Chris_Pine}).
- $e_{12}$ = ($v_1$, $v_2$, *Association*, FILMEDIN, −), $e_{14}$ = ($v_1$, $v_3$, *Association*, DIRECTS, -), $e_{15}$ = ($v_1$, $v_5$, *Association*, ACTS, {ranking}), $e_{15}$ = ($v_1$, $v_6$, *Association*, ACTS, {ranking}), $e_{23}$ = ($v_2$, $v_3$, *Composition*, LOCATEDIN, −).

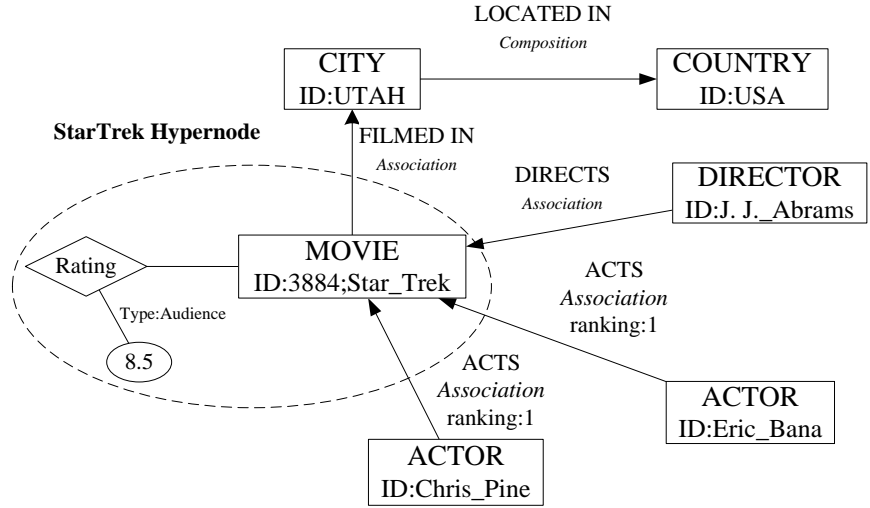

Figure 4. *MovieLens* with GRAD notations

## 5. INTEGRITY CONSTRAINTS

Integrity constraints are the general rules describing the consistent database states, or change of states, or both (Codd, 1980). They play a fundamental role in data quality enforcement within data management systems, and especially for decision support systems. Integrity constraints were first introduced with the relational model and later studied in object-oriented and semi-structured data models such as XML. This topic is also present in the context of graph databases, where the efforts were centered around applying integrity constraint concepts from the relational model on graph data (Angles, 2012). In this section we provide a formal definition for the integrity rules applied over the graph structures presented in the previous section.

As discussed in Angles et al. (Angles & Gutierrez, 2008; Angles, 2012), several integrity constraints were proposed for graph databases in the literature. We identify the ones relevant to GRAD as follows:

- *Graph entity integrity*: This constraint is used to guarantee that each real-world entity is represented by a unique hypernode. It also provides the mechanisms for nodes and edges identification through specific attributes (i.e., ID) and/or structural properties such as neighborhoods.
- *Semantic constraints*: These are user-defined constraints. The first type represents the assertions (i.e., topological and value-based constraints) users wish to define on the graph elements. The second type focuses on cardinality checking between classes of nodes. We now study both kinds of constraints in more detail.

### 5.1. Graph Entity Integrity

This first category of integrity constraints aims at guaranteeing that a real-world entity is represented only once in the graph database. This prevents data redundancy of data and helps to fulfill consistent update and deletion of graph entities. Since we are managing labeled graphs, we need to ensure a set of constraints on the labeling of graph elements. These constraints are defined as follows:

- Each class should have exactly one unique label different from all the labels of the other classes: $\forall \Sigma_i, \Sigma_j \in \sum, \text{if } \Sigma_i \neq \Sigma_j \text{ then } C_i \neq C_j$.
- Entity edge labels between different classes are unique: let $v_i, v_j, v_k \in V_e$, $\mathcal{F}_e(v_i, v_j) = \mathcal{F}_e(v_i, v_k)$ iff $\mathcal{F}_v(v_j) = \mathcal{F}_v(v_k)$.
- Two entity edges with the same label cannot link the same pair of entity nodes: let $v_i, v_j \in V_e$ and $l_1, l_2 \in \mathcal{L}_e$, if $e_m = (v_i, v_j, \text{type}_m, l_1, B_m)$ and $e_n = (v_i, v_j, \text{type}_n, l_2, B_n)$ then $l_1 \neq l_2$.

Entities in the graph are identified by their unique identifier, their neighborhood or both, using the following identification mechanisms:

- Each entity node should have an identifier that is unique within the class of that node. Therefore each entity node $v_i$ is uniquely identified by the pair composed of its identifier ($ID_i$) and its label ($C_i$): $< C_i , ID_i>$. The exception here is on the identification of weak entity nodes (i.e., entity nodes related by a composition relationship to another composite entity node). These weak entity nodes require also the identifier of their parent entity node to be identified. Let $ID_j$ be the identifier of a weak entity node $v_j$, whose parent is identified by $ID_{parent}$; then $v_j$ is identified by the triple $<C_j, ID_{parent}, ID_j>$.
- GRAD supports multigraphs. Thus to unequivocally identify an entity edge, the identifiers of the nodes it links are not enough, we also need the edge's label. We previously stated that two entity edges with the same label cannot link the same pair of entity nodes. Therefore, each entity edge $e_i$ is uniquely identified by the triple comprised of its label ($l_i \in \mathcal{L}_e$) and the identifiers ($ID_j, ID_k$) of the entity nodes it links ($v_j$ and $v_k$): $<l_i, ID_j, ID_k>$.
- An attribute node $v_i \in V_a$, associated to an entity node $u \in V_e$ is identified by the pair comprised of its label ($l_i \in L_a$) and the identifier ($ID_j$) of u: $<l_i, ID_j>$. An attribute edge is identified by the entity node and attribute node it links.

### 5.2. Semantic Constraints

Based on the knowledge of a specific domain, end-users might need to make some semantic restrictions over the graph data. The goal of these constraints is to guarantee the compliance of the graph data with respect to a given domain specific rules. It is clear that these constraints could not be automatically identified and captured by the system. They need to be explicitly expressed by users. In GRAD, we also choose them to be represented using graph patterns. Thus, prior to defining the notion of semantic constraints, we must introduce the concept of *graph pattern in GRAD*. A graph pattern P is a predicate on the topology (specifying conditions on the structural properties of the graph) and attributes (specifying conditions on the values of the attributes) of the graph elements. The topic of graph pattern matching is well-studied in the graph theory literature (Fan, 2012), we formally define a graph pattern in GRAD as follows:

**Definition 11** (Graph Pattern) A ***graph pattern*** is defined as $\mathcal{P} = (V_p, E_p, \alpha_v, \alpha_l, \alpha_e, \beta_v, \beta_e)$, where:

- $V_p$ is a finite set of nodes, with $V_p \subseteq \mathcal{V}$.

- $E_p$ is a finite set of edges, with $E_p \subseteq \mathcal{E}$.

- $\alpha_v$: is a function defined on $V_e$ such that for a given node $u \in V_e$, $\alpha_v(u)$ is the predicate applied on the identifier $ID_i$ of $u$. This predicate is a conjunction of atomic predicates such that each predicate compares a constant $c$ specified in the pattern with the value $ID_i$ using a given operator $op$. The comparison is performed using any of the following operators: $<, \leq, =, \geq, >, \neq$. Let $c_j$ be a constant and $op_j$ be a comparison operator, $\alpha_v(u)$ is the conjunction of atomic predicates of the form: $ID_i \ op_j \ c_j$.

- $\alpha_l$: is a function defined on $V_l$ such that for each node $u \in V_l$, $\alpha_l(u)$ is the predicate applied on the value *val* stored in the literal node $u$. This predicate is a conjunction of atomic formulas such that each of them compares a constant $c$ specified in the pattern with the value *val* using a given operator $op$. The comparison is performed using any of the following operators: $<, \leq, =, \geq, >, \neq$. Let $c_j$ be a constant and $op_j$ be a comparison operator, $\alpha_l(u)$ is the conjunction of atomic predicates of the form: *val* $op_j \ c_j$.

- $\alpha_e$: is a function defined on $E_e \cup E_l$ such that for each edge $e \in E_e \cup E_l$, $\alpha_e(e)$ is the predicate applied on the attributes of $e$. This predicate is a conjunction of atomic formulas that each of them compares a constant $c$ specified on the pattern with the actual value of an attribute of the edge $e$ using a given comparison operator $op$. The comparison is performed using any of the following operators: $<, \leq, =, \geq, >, \neq$.

- $\beta_v$ is a function defined on $V_e \cup V_a$ such that for each node $u \in V_e \cup V_a$, $\beta_v(u)$ is the predicate applied on the label of $u$ (i.e., the predicate is applied on $C_i$ if $u \in V_e$, and on $l_i$ if $u \in V_a$). This predicate is used to compare the string specified on the pattern with the actual label of the node. Given a label $C_i$ of $u$ and a string $s_i$, the comparison is of the form $C_i \ op \ s_i$, and is performed using one of the two equality comparison operators $=, \neq$.

- $\beta_e$ is a function defined on $E_e$ such that for each node $e \in E_e$, $\beta_e(e)$ is the predicate applied on the label of $e$. This predicate is used to compare the string specified on the pattern with the actual label of the edge. Given a label $l_i$ and a string $s_i$, the comparison is of the form $l_i \ op \ s_i$, and is performed using one of the two equality comparison operators $=, \neq$.

We represent each atomic formula of the predicate functions (i.e., $\alpha_v, \alpha_l, \alpha_e, \beta_v, \beta_e$) as $f_{i(op_j, c_j)}(val) \rightarrow \{true, false\}$. The predicate function is evaluated to true if all its atomic formulas are true. In this paper, we focus on the case of conjunction between predicates. However, the same approach could be extended to support disjunctions or an arbitrary combination of conjunctions and disjunction of predicates.

In what follows, we use subgraph isomorphism to identify all the subgraphs that match a given graph pattern. We say that a subgraph G' of a given graph G matches a query pattern $\mathcal{P}$ by isomorphism, and we denote this as: G' $\cong \mathcal{P}$.

**Definition 12** (Subgraph Matching) Consider a GRAD graph $\mathcal{G}$. A subgraph of $\mathcal{G}$, denoted G' = (V', E', $\mathcal{L}_v$, $\mathcal{F}_v$, $\mathcal{L}_e$, $\mathcal{F}_e$, $\Lambda_v$, $\mathcal{H}_v$, $h_l$, $\Lambda_e$, $\mathcal{H}_e$), *matches* a pattern $\mathcal{P}$ using graph isomorphism, if there is a bijective function $\Phi: V_p \to V'$ such that:

- Every node on $V_p$ has an image node on V' by the bijective function $\Phi$, and every edge from $E_p$ has an image edge on E'. Formally, $\forall u, u' \in V_p, e = (u, u') \in E_p$, iff $\Phi(u), \Phi(u') \in V'$ and $(\Phi(u), \Phi(u')) \in E'$.

- $\forall u \in V_e'$, $\alpha_v(u)$ holds, i.e., $\forall f_i$ such that $f_i$ is an atomic predicate of $\alpha_v$ where $f_i$ applies on the attribute $a^i = h_v^i(u)$: $f_{i(op_j, c_j)}\left(h_v^i(u)\right) = true$.

- $\forall u \in V_l'$, where u is linked to its attribute node v through the literal edge e such that e = (v, u), $\alpha_l(u)$ holds, i.e., $\forall f_i$ such that $f_i$ is an atomic predicate of $\alpha_l$ where $f_i$ applies on the value $val = h_l(v, e)$: $f_{i(op_j, c_j)}\left(h_l(v, e)\right) = true$.

- $\forall e \in E_e' \cup E_l'$, $\alpha_e(e)$ holds, i.e., $\forall f_i$ such that $f_i$ is an atomic predicate of $\alpha_e$ where $f_i$ applies on the value $b^i = h_e^i(e)$: $f_{i(op_j, c_j)}\left(h_e^i(e)\right) = true$.

- $\forall u \in V_e' \cup V_a'$, $\beta_v(u)$ holds, i.e., $\forall f_i$ such that $f_i$ is an atomic predicate of $\beta_v$ where $f_i$ applies on the label $l_i = \mathcal{F}_v(u)$: $f_{i(op_j, c_j)}(\mathcal{F}_v(u)) = true$.

- $\forall e \in E_e'$, $\beta_e(e)$ holds, i.e., $\forall f_i$ such that $f_i$ is an atomic predicate of $\beta_e$ where $f_i$ applies on the label $l_i = \mathcal{F}_e(e)$: $f_{i(op_j, c_j)}(\mathcal{F}_e(e)) = true$.

It should be noted that the bijective function $\Phi$ specifies the minimum set of properties a graph element should have. For a pair of nodes $v \in V_p$, $v_x \in V'$, $v_x = \Phi(v)$ means that (1) $v_x$ has at least all the attributes of v for which the predicates of $\mathcal{P}$ hold, and (2) $v_x$ has at least all the relationships v has. However, the matched nodes ($v_x \in V'$) could have more relationships and attributes than those specified on the pattern's description.

The problem of graph matching using subgraph isomorphism is known to be NP-complete. Indexing techniques were proposed in literature to accelerate matching algorithms. For some domains and applications, weaker forms of graph matching are sufficient. Hence, the strict constraint of isomorphism is relaxed, reducing the complexity of the matching operation. Among these alternatives, we can cite graph pattern matching by simulation, dual simulation, strong simulation, and bounded simulation (Fan, 2012).

We are now ready to address semantic constraints. In particular, we study here two categories of such constraints: assertions and multiplicities.

*Assertions* These are predicates applied on the graph data and that must always be satisfied. In GRAD, assertions are represented using graph patterns to specify the topological and content-based constraints chosen by the user.

**Definition 13** (Assertion) An assertion is represented by a graph pattern $\mathcal{P}$. A graph G is said to be compliant to the assertion if for every entity node ($v \in V_e$) matched by a node on the assertion pattern ($v_x \in V_p$), there exist a subgraph G' ⊆ G such that G' matches $\mathcal{P}$.

Formally: $\forall v_e \in V_e, v_x \in \mathcal{P} \mid \Phi(v_x) = v_e \Rightarrow \exists! \ G' \subseteq G \mid v_e \in G'$ and $G' \cong \mathcal{P}$.

**Example 9** (Assertion on Movies) For example, a user can state that all movies on the graph should have an attribute node labeled "*Rating*", with the rating value by "*Audience*" being above 7. The movie has to be related to a set of actors and a director. If these conditions are not met, the insertion transaction of the movie in the graph should fail. The pattern predicate is represented graphically in Figure 5 (i.e., as a graph whose structure represents the topological constraints and the values on its attributes represent the conditions on the output attributes). This example pattern is represented in GRAD as follows:

- Set of types of nodes: $V_e = \{v_A, v_M, v_D\}$, $V_a = \{v_r\}$ and $V_l = \{v_l\}$.
- Set of types of edges and their respective end-nodes: $E_e = E_{as} = \{e_D, e_A\}$, with $e_D = (v_D, v_M)$ and $e_A = (v_A, v_M)$. $E_a = \{e_r = (v_M, v_r)\}$ and $E_l = \{e_t = (v_r, v_l)\}$.
- Predicates on the labels of entity nodes: $\beta_v(v_A): C_{v_A} = $ "ACTOR", $\beta_v(v_M): C_{v_M} = $ "MOVIE" and $\beta_v(v_D)\ C_{v_D} = $ "DIRECTOR".
- Predicate on the label of the attribute nodes: $\beta_v(v_r): l_{v_r} = $ "Rating".
- Predicates on the labels of entity edges: $\beta_e(e_A): l_{e_A} = $ "ACTS" and $\beta_e(e_D): l_{e_D} = $ "DIRECTS".
- Predicate on the attribute of literal edges: $\alpha_e(e_t): h_e^{Type}(e_t) = $ "Audience".
- Predicate on the value of literal nodes: $\alpha_l: h_i(v_r, e_t) > 7$.

*Multiplicities* They are applied between classes of entity nodes to define the number of relationships a node from a given class can have with nodes from other classes through a particular entity edge. This is a user-defined constraint that is well-known in conceptual modeling languages. In GRAD, multiplicities specify the maximum and minimum number of relationships an entity node is allowed to have with entity nodes from other classes, through entity edges of a given label. If not set by the user, the default multiplicity is unbounded.

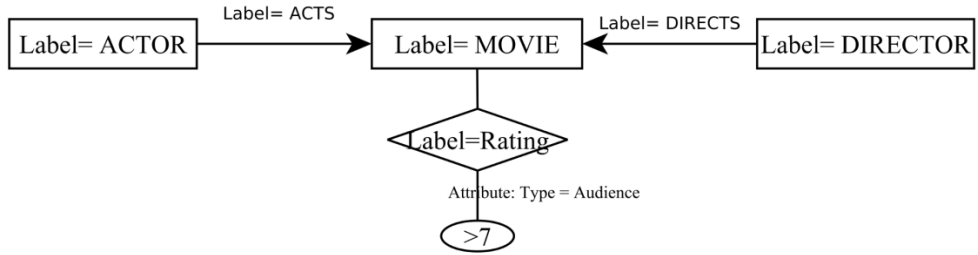

Figure 5. Assertion pattern on movie nodes

**Definition** 14 (Multiplicity) Formally, this constraint is represented by a function $\mathcal{M}$ defined on $\Sigma \times L_e \times \Sigma$. For a given pair of entity node classes ($\Sigma_i, \Sigma_j \in \Sigma$), and an edge's label ($l_k \in L_e$), $\mathcal{M}(\Sigma_i, l_k, \Sigma_j)$ defines the multiplicity predicate. It specifies the maximum and minimum number of entity edges labeled $l_k$ that could exist between a given node $v_i \in \Sigma_i$ and the set of nodes $v_j \in \Sigma_j$, and vice versa.

Note that the multiplicity predicate is a conjunction of at most two atomic formulas (specifying the lower and upper bounds) at each endpoint. Each predicate compares an integer $i$ with the number of edges using a given comparison operator from the following list: $<, \leq, =, \geq, >, \neq$.

**Example 10** (Multiplicity of Movies) A user may want to state that an actor can participate in many movies but that a movie should have at least one actor. In the first case, the maximum range is unbounded, while the second range is greater than one. We represent this constraint between actors and movies related by the *ACTS* relationship using UML notations as [1..*, *]. Figure 6 illustrates this example of constraints using a graph-based representation. Note that the multiplicities should not violate the specific properties of entity edges (e.g., to-one relationships between nodes related by composition).

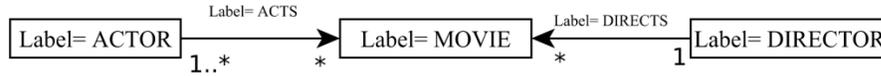

Figure 6. Multiplicity of movie nodes

While multiplicities focus solely on cardinality checking between classes of nodes, assertions extend the constraints' scope to target the topology and the content allowed on the graph. Naturally, the two constraints could be combined by adding for example the function M to the assertion's pattern 𝒫. In such case, the assertions will define the constraints on the topology and attributes of graph elements, and the multiplicity defines the allowed number of relationships between the different entity nodes.

Many other integrity constraints were studied on the literature of graph database models. For example, functional dependencies and referential integrity were considered by some graph database models. Functional dependencies describe the fact that values of an element's attribute determine the values of another one. Referential integrity, on the other hand, as defined on the relational model, guarantees that only existing entities are referenced. These two constraints are designed to put constraints on implicit relationships existing between data entities. However, these relationships are explicitly expressed on graph using edges. We can therefore use specific graph edges (e.g., composition) to reflect the same semantic. Moreover, as acknowledged in (Angles & Gutierrez, 2008), these concepts are inherited from the relational model and remain difficult to project directly on the graph database models.

## 6. GRAPH ALGEBRA

In this section, we complete the GRAD database model with a graph algebra consisting in a set of algebraic operators specific for online graph querying and analysis. The algebra supports the data structures of GRAD while preserving the integrity constraints. This allows users to traverse and query the network without violating the database integrity. The algebra we present in this section extends GraphQL, a graph algebra defined along the lines of the relational algebra (He & Singh, 2010). We discuss later our choice of GraphQL as the basis of our algebra and its positioning with regard to other graph algebras. GRAD algebra operates directly on GRAD structures, which are the fundamental construct, the operands and the return type of all algebraic operations. By building on GraphQL, our algebra inherits the same expressive power; hence it is at least as expressive as the relational algebra. We study the closure of each algebraic operator with respect to GRAD structures and integrity constraints. When starting from a consistent database state (i.e., where the integrity constraints presented above are satisfied), the closure ensures that the integrity constraints remain satisfied after the execution of any given algebraic operation. This guarantees that each operator's output is formatted according to GRAD and thereof be a valid input for the other operators. In other words, all algebraic operators could be pipelined and the result is a graph compliant to the integrity constraints defined above. Finally, algebraic laws regarding commutativity and associativity remain valid since the algebra is built on the lines of relational algebra.

For all the operators we introduce in this section, we assume the following:

- The input is a GRAD graph.

- The general rule stating that if a node is removed, then all its edges are removed, is valid for all types of nodes.

- If a composite (parent) entity node is removed, all its parts (children) weak entity nodes are removed consequently. This means, for instance, that if an entity node is removed then the whole hypernode is removed, and if an attribute node is removed then all its literal nodes are removed as well.

We now study the operators one by one.

## 6.1. Selection

The first major operation is the selection which is a subgraph extraction operation based on graph pattern matching. Selection consists in finding the subgraph (or set of subgraphs) of the data graph $\mathcal{G}$ that satisfies the semantic and topological constraints specified by a given graph pattern $\mathcal{P}$.

The selection operates only on valid graph pattern in GRAD defined as follows:

**Definition** 15 (Valid Pattern) A given graph pattern $\mathcal{P} = (V_p, V_e, \alpha_v, \alpha_l, \alpha_e, \beta_v, \beta_e)$ is a valid GRAD pattern if it satisfies the following conditions:

- All elements of $\mathcal{P}$ are represented using GRAD structures: $V_p \subseteq V_e \cup V_a \cup V_l$ and $E_p \subseteq E_e \cup E_a \cup E_l$. Figure 7(a) shows an example of a subgraph that violates this condition. The violation here is that two attribute nodes are linked to the same literal node while in GRAD a literal node must be related to exactly a single attribute node.

- The definition of predicates on the content (label and attributes) of the pattern elements is optional. In case no predicate is specified on an attribute or a label, all elements match the pattern. Intuitively, for a given graph element, a predicate can only be applied on the properties specified for its kind by GRAD (e.g., a predicate cannot be applied on the labels of attribute edges because by definition an attribute edge do not have a label). For example the pattern on Figure 7(a) states that the attribute edge linking an actor entity node to its ranking attribute node should be labeled "Rank", while GRAD does not support labels on attribute edges.

- The pattern does not include weak nodes without their parent nodes. That means, if the pattern includes a weak node (i.e., child nodes related by a composition relationship to their parent nodes), then the parent node should be included in the pattern as well. This is required because a weak node does not represent a complete entity and could not be identified by itself. For example if the pattern describes an attribute node and its literal node, then it should also contain a description of their corresponding parent entity node. For example Figure 7(b) is not valid because the attribute node is not associated with an entity node. Figure 7(c) is not valid because entity nodes labeled "CITY" are weak entity nodes linked by composition to their parent entity nodes labeled "COUNTRY" which are not included in the pattern.

Regarding the *MovieLens* working example, Figure 5 and Figure 8 provide examples of valid GRAD patterns. Particularly, Figure 8(a) and Figure 8(b) provide a possible correction for the non-valid patterns shown in Figure 7(b) and 7(c) respectively.

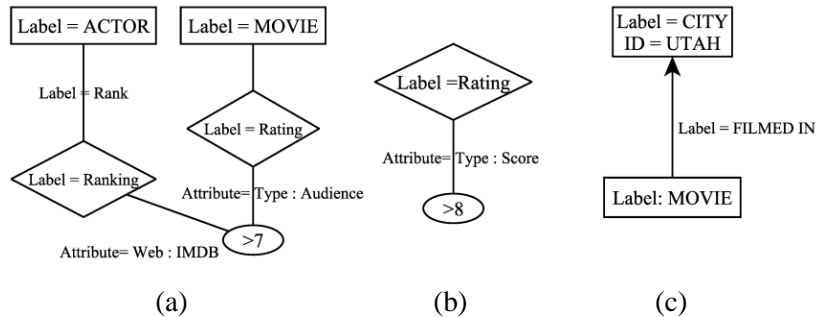

(a)    (b)    (c)

Figure 7. Non-Valid GRAD patterns

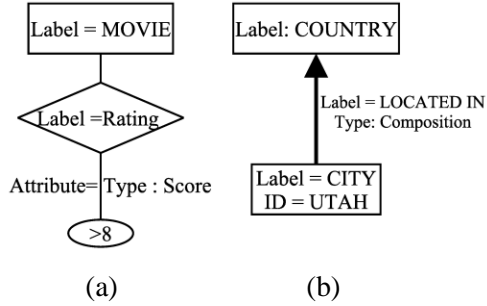

(a)         (b)

Figure 8. Valid GRAD patterns

Using the formal definitions of GRAD data structures and valid GRAD pattern, we formally define the *selection operator* as follows:

- Input: A graph $\mathcal{G}$ and a selection pattern $\mathcal{P}$.

- Mathematical notation: $\sigma_{\mathcal{P}}(\mathcal{G})$.

- Output: A set of matched graphs $\Omega = \{(G_1, G_2, \ldots, G_n) | G_i \cong \mathcal{P}\}$, such that all $G_i$ are subgraphs of $\mathcal{G}$ that match the selection pattern $\mathcal{P}$ by isomorphism.

- Closure condition: The closure is satisfied if the selection pattern is a valid GRAD pattern.

**Example 11** (Selection operation) Suppose the analyst would like to extract the network of the top-rated movies by audience (i.e., movies rated above 7 by the audience). The user is also interested in getting the top actors of each top movie (i.e., actors with ranking equal to 1 playing in the selected top-rated movies). This query is answered by applying the following selection: $\sigma_{\mathcal{P}}(\mathcal{G}_{MovieLens})$. Here the pattern applies constraints on the labels of entity nodes ("*ACTOR*" and "*MOVIE*"), the label and attribute of their linking entity edge (label is "*ACTS*", and attribute is "*Ranking = 1*"), the labels of movies' attribute nodes (label is "*Rating*") and the rating values (value in the literal node > 7). The input and output of this selection operation are shown on Figure 9.

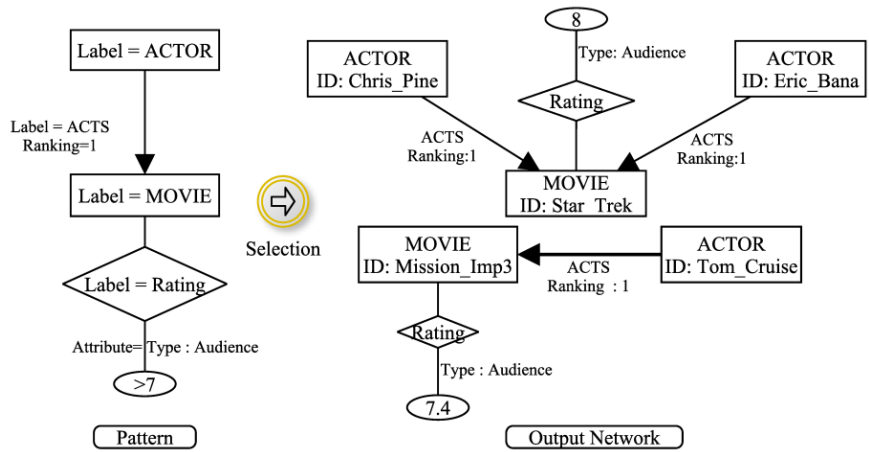

Figure 9. Top movies and actors network

Note that even if the pattern is not GRAD compliant, the selection could be performed to extract graph data. However, the result cannot be guaranteed to be a valid input for other operations as it violates their preconditions.

## 6.2. Cartesian Product

*Cartesian Product* is a binary operator applied to put together two collections of graphs. The algebra of this operator and is defined as follows:

- Input: Two collections of graphs $S_1$ and $S_2$.

- Mathematical notation: $S_1 \times S_2$.

- Output: Let $S_1 = \{G_1, G_2\}$ and $S_2 = \{G_3, G_4\}$. The result of the Cartesian product is a set of pairs of unconnected graphs. Each graph on the output is composed by a graph from $S_1$ and a graph from $S_2$ respectively: $S_1 \times S_2 = \{G_a = (G_1, G_3), G_b = (G_1, G_4), G_c = (G_2, G_3), G_d = (G_2, G_4)\}$.

- Closure condition: Cartesian product is a closed operation since it does not alter the internal structure of the hypernode neither adds new edges.

## 6.3. Composition

The idea of *composition* is to reuse the information extracted from the input graph data in order to *generate* new graphs. The composition operator is used to create a new graph based on data collected from the original graph, then formatted according to a given graph template. Basically, a graph template $\tau$ describes the structure of the output graph by specifying the content of graph entities and their organization on the result graph. The template could be defined using a subgraph G'. As shown in Figure 10, the template describes an output network with one type of entity nodes (labeled "*ACTOR*" and with the identifier "name"). Entity nodes are related by one type of relationships (labeled "*Co-Acts*" and of type "Association").

Applying composition is to some extent analogous to having a function with predefined instructions (i.e., a graph template), a set of formal parameters (i.e., a graph pattern), and a set of actual parameters values (i.e., the actually matched graphs). First, the user needs to define the graph template that she wants to generate using data from the input graph. Then a graph pattern matching operation is applied on the input graph to retrieve a set of matched subgraphs. Finally, the composition operator instantiates the graph template using data from the matched subgraphs. The composition operator is formally defined as follows:

- Input: The initial graph $\mathcal{G}$, the graph template $\tau$, and a pattern $\mathcal{P}$.

- Mathematical notation: $\omega_{\tau_p}(\mathcal{G})$.

- Output: Let $\Omega$ be the set of graphs matched by the pattern $\mathcal{P}$ such that $\Omega = \sigma_p(G)$. Let $\tau_p(G)$ be the function that instantiates the template $\tau$ given data matched by $\mathcal{P}$ on an actual graph $\mathcal{G}$. The output of the composition is denoted as $\omega_{\tau_p}(G) = \{\tau_p(G_i) \mid G_i \in \Omega\}$. Hence, the result is a set of graphs instantiating the template $\tau_p$ (i.e., graph elements are created according to the template and the actual values of their attributes are assigned).

- Closure condition: The composition is closed if the template is compliant to GRAD structures and constraints.

**Example 12** (*Composition operation*) Assume the analyst wants to generate the graph of co-actors from the original graph. The template is a network of actors connected by co-actorship edges. The actual names of the actor are retrieved from the input graph. The pattern matches the pairs of actor nodes that are directly connected to the same movie. This operation is depicted in Figure 10.

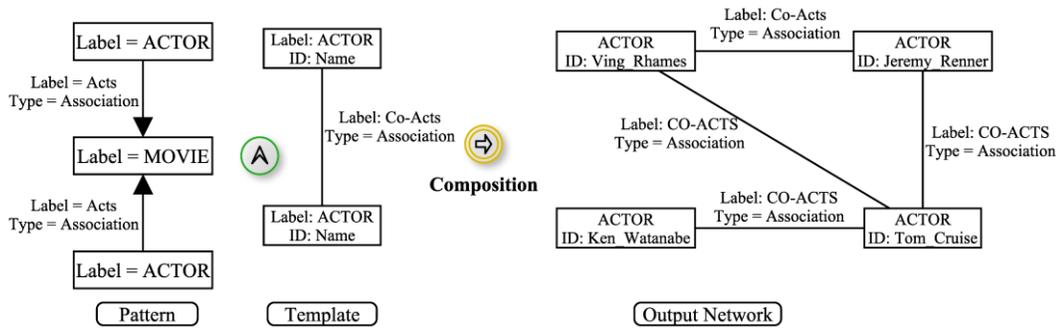

Figure 10. Co-actorship network generation

### 6.4. Set operators: Union and Difference

The two core set operations on graphs are union and difference. A union between two (collections) of graphs simply generates a new graph putting together the two input graphs without concatenation. Union does not introduce changes to the internal structure of each input graph. Graph union is different from its relational counterpart in that no common structure is required for executing the operation.

The *union operator* is defined as follows:

- Input: Two initial graphs G and G'.

- Mathematical notation: G ∪ G'.

- Output: G ∪ G' = G'', where all graph elements from G and G' are on G'' (i.e., $\mathcal{V}'' = \mathcal{V} \cup \mathcal{V}'$, and $\mathcal{E}'' = \mathcal{E} \cup \mathcal{E}'$).

- Closure condition: The union does not introduce any structural changes to the graph elements. This operation is therefore closed.

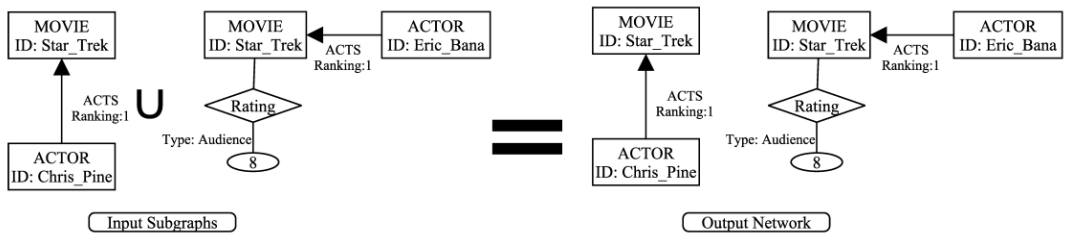

Figure 11. Union of actors of the same movie

**Example 14** (Graph Union): While adding new graph sources to the graph database, the union operation is used simply to put together the original graph with the graph entities being added. The union is sufficient if the added entities do not exist previously on the graph. Note that as in Figure 11 the two subgraphs are unified but not merged. Intuitively, if an entity (e.g., the *Star Trek* movie) is already present in the graph, an integration operation fusing the two corresponding entity node is required. We further investigate this case with the join operation introduced next.

The difference between graphs removes isomorphic subgraphs that exist in the two input graphs. The *Difference operator* is defined as follows:

- Input: Two graphs G and G'.

- Mathematical notation: G – G'.

- Output: G – G' = G", where all graph elements from G isomorphic to graph elements in G' are removed from G. While composition is used to generate new graphs, difference is the only operation we introduced so far to enable deletion of graph elements.

- Closure condition: The difference introduces structural changes to the graph by removing the shared subgraphs. This operation is closed as long as it respects the assumptions we introduced at the beginning of this section.

## 6.5. Structural Graph Join

The union or Cartesian Product operations introduced above enable putting together a collection of graphs. However in scenarios where graph entities need to be integrated these operation are not enough to merge redundant graph entities. We need therefore to introduce another operation, the ***structural graph join***, or simply join, which consists in the unification of nodes and edges based on a common join predicate. The join operator is defined as follows:

- Input: Two collections of graphs $S_1$ and $S_2$, and a join predicate *Pr*.

- Mathematical notation: $S_1 \bowtie_{Pr} S_2$.

- Output: A join is performed in multiple steps. First a Cartesian Product: $S_1 \times S_2$ is applied and the result is a set of pairs of subgraphs. For this set of pairs, a selection is applied to retain only pairs of subgraphs whose elements satisfy the join predicate *Pr*. For each retained pair of subgraphs, their graph elements that satisfy the predicate are merged. The generation of the new graph resulting from the join is performed by the composition operation. However, this is a specific case of composition where there is no need for the pattern matching phase since the graphs to join are the input instead of the matched subgraphs. For the composition template, it consists in the union of the two joined graphs after edges and nodes unification.

- Closure condition: The join operation is closed in case no unification is required. In case of unification of edges or nodes, the structure of the graph might change and the closure is verified at the level of the composition operation that actually performs the unification.

**Example 13** (Structural Join): Assume the analyst wants to integrate the incoming data about movies in the graph already stored in the database. Often, some movies and actors could have been already inserted in the database. These existing entities have to be merged with their corresponding incoming ones. In this example, the correspondence between the existing nodes and potential incoming nodes is performed based on the identifiers of the movie nodes. After the fusion, we can notice that the two movie nodes (with *ID*= 3884) are merged and an additional type of attribute nodes (*Score*) is attached to the existing movie as shown on the first join on Figure 12. Once the entity nodes and their attribute nodes are merged, another possible scenario could be the fusion of attribute nodes of the movie node. This is shown in the second join on Figure 12, this adding the new captured rating values as literal nodes attached to the *Rating* attribute node.

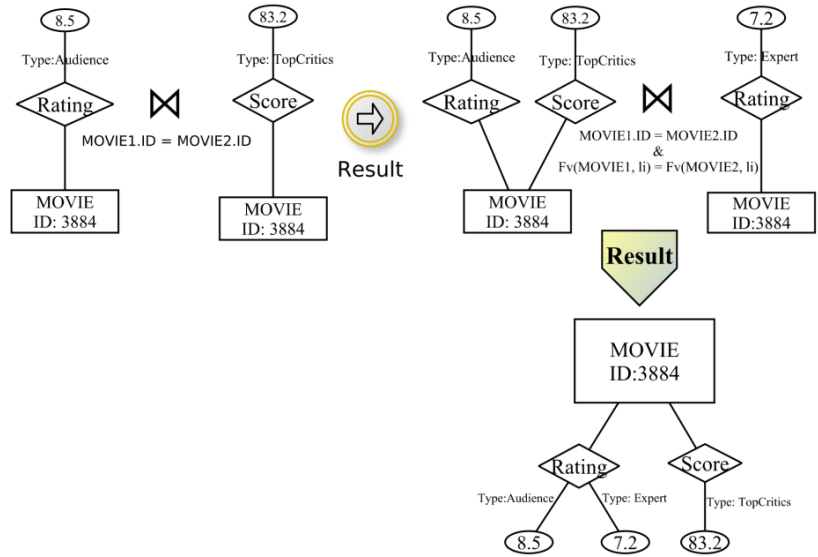

Figure 12. Join on attribute and value nodes

## 7. APPLICATION FIELDS

Graphs provide a generic and powerful abstraction tool for knowledge representation. Graph-based data models are more generic than relational tables or XML documents. Many widely-used models are represented using graph structures. For example, the entity-relationship model (Chen, 1976) is represented as a set of nodes describing entities related with edges representing different relationships between them. Ontologies are also represented as graphs of inter-related primitive entities from the same domain. XML files are structured as trees, which could be considered as specific graphs limited to the hierarchical model rather than the networked model. GRAD provides a more advanced abstraction than primitive property graphs, while remaining generic enough to model different applications. Here we show two such potential applications of GRAD.

*Master Data Management* Data integration is a critical task in multiple data management systems. GRAD equips analysts with powerful operators enabling the integration process while ensuring the consistency of the database. Lim et al. (2010) investigated techniques for data integration using graphs at both the instance and schema levels. Particularly, Master Data Management (MDM) tools are a potential candidate for the use of GRAD. Master data represent the set of critical entities shared across an organization's infrastructure. It evolves with the introduction of new data sources, for instance through the merging or acquisition of organizations. MDM tools are used in industry as a middleware to integrate multiple data management systems and provide a unified view of the master data. GRAD graphs are flexible and foster data integration using the concept of hypernode and the uniqueness integrity constraints. Each master data element can be represented using a unique hypernode. Usually, in data integration scenarios, data flowing in graph data management systems either include new real-world entities, or additional information about existing entities. With the integration of new data management systems in the MDM, data belonging to existing entities in the MDM can be seamlessly integrated in its corresponding hypernode. Otherwise, a new hypernode will be created. Moreover, GRAD supports various types of relationships that can further help to capture more semantics about the master data. Finally, graph-oriented queries, such as pattern discovery and paths or neighborhood exploration, could be used for advanced master data analysis.

As an example of the above discussion, in the case of the *MovieLens* network, movies and actors could be considered to be part of the master data. Initially, data about movies are stored in many formats across various data stores. At the MDM level, each movie can be represented by a movie hypernode. The overall

MDM graph changes in content and structure with the insertion of new classes such as *USER* or new connections between new movies and existing actors etc. We capture this graph evolution by adding new classes of entity nodes with their respective entity edges and attributes. Clearly, graph-based data models like GRAD are appropriate to easily handle such changes.

*Data Warehouses* Data warehouses (DW) integrate data from many sources in an organization, to allow data analysis and mining. Traditional data warehouses are built using the multidimensional model, usually implemented over a relational database, like in the case of ROLAP -Relational OLAP - systems). They provide a subject-oriented, historical and integrated view of the data. This makes data warehouses a suitable backbone for common analysis techniques such as Online Analytical Processing (OLAP), reporting and data mining. DWs are thus a key component of decision-support systems (DSS). The first task in a data warehouse system is the Extract-Transform-Load process (ETL), which integrates data from the sources in order to populate the data warehouse. As discussed earlier, GRAD fosters data integration. Hence, it could serve at both the ETL and the multidimensional design phases. In a graph setting, the ETL process should support tasks such as identification, transformation, matching and insertion of the incoming nodes and edges in the graph data warehouse. As explained through the paper, the algebra, and specifically the composition and join operations are particularly well-suited for the implementation of the transformation phase. Moreover, data quality could be enforced using integrity constraints checking mechanisms. At the multidimensional schema design phase, each class of hypernode is a potential dimension. Multiplicities are implicit for specific types of edges (e.g., composition and aggregation are to-one relationships). The information inferred form relationship types or explicitly given as a constraint by the user could help in designing the OLAP dimension hierarchies, which support the data aggregation process, which is crucial in an OLAP setting. We do not extend in OLAP and data warehouse concepts here since the discussion is beyond the scope of this work but we refer readers to Li et al. (2011), Ghrab et al. (2013), and Wang et al. (2014) works for examples of graph data warehousing frameworks.

## 8. SYSTEM IMPLEMENTATION

As a proof-of-concept of the applicability of GRAD, we present here the modular architecture of a graph management system designed using GRAD principles. The *GRAD CORE* is the central library implementing the concepts discussed throughout the paper. It is domain and storage independent. The architecture of the system is depicted in Figure 13. Here, the design goal is to extend traditional DSSs to graph data using GRAD. The system consists of four major components defined as follows:

- *GRAD ETL*: Graph data are first extracted from external data sources. The input data might be under various formats (e.g., XML as for DBLP, or text data files for *MovieLens,* etc.). This first step consists mainly in the development of a parser tailored to GRAD in order to transform and integrate input data into a GRAD-compliant graph store. This phase is similar to the traditional ETL process in data warehouse systems. One parser per data source is required at this phase. The GRAD algebra (and specifically, composition and join operations) plays a major role during this phase to integrate and match incoming data, while preserving the integrity constraints. In our implementation, we used the *MovieLens* dataset as input. The ETL is done by *MovieLens2GRAD*, which is a parser extracting the graph data from multiple files, and integrating them into the storage back-end.

- *Storage Back-end*: At this level a database engine is required for efficient storage and retrieval of data. The graph data is persisted on disk and accessed using the underlying API of the storage system. In our example, the *GRAD4Neo* module was developed as a wrapper around the Neo4j graph database. It receives as input the graph data from *MovieLens2GRAD* and inserts it into the datastore using the core API of Neo4j. *MovieLens2GRAD* is optimized and tailored to reuse the advanced capabilities of Neo4j to efficiently store, index, and query graph structures. GRAD is a database model, therefore a discussion on optimization at the physical level is beyond the scope of this paper.

Note that although we chose Neo4j for its performance and rich API, the same approach could be applied to other datastores (e.g., Triple stores or relational databases). A wrapper per datastore needs to be developed at this level.

- *GRAD CORE:* GRAD was designed with modular architecture that makes it independent from the data sources and the physical storage back-end. *GRAD CORE* is the central library that implements the GRAD data structures, integrity constraints and algebra presented along the paper. It is written in Java and defines the set of interfaces that need later to be implemented and tailored by the wrapper to the specific API of each database engine. The algorithmic library that implements the algebraic operator and preserves the integrity constraints is also part of this component. This component implements the core of our database model and is domain and datastore independent.

- *Graph Operations:* The graph querying API and the graph visualization web interface are designed to simplify users' effort in querying graph data. They are built on top of the *GRAD Core* module. GRAD could be easily extended with graph analytics libraries. Such libraries would contain graph algorithms (written and optimized in advance) for analyzing and computing common measures on top of graph data (such as the shortest path, Page Rank and eigenvector centrality). Using these libraries, rich graph analysis scenarios could be achieved.

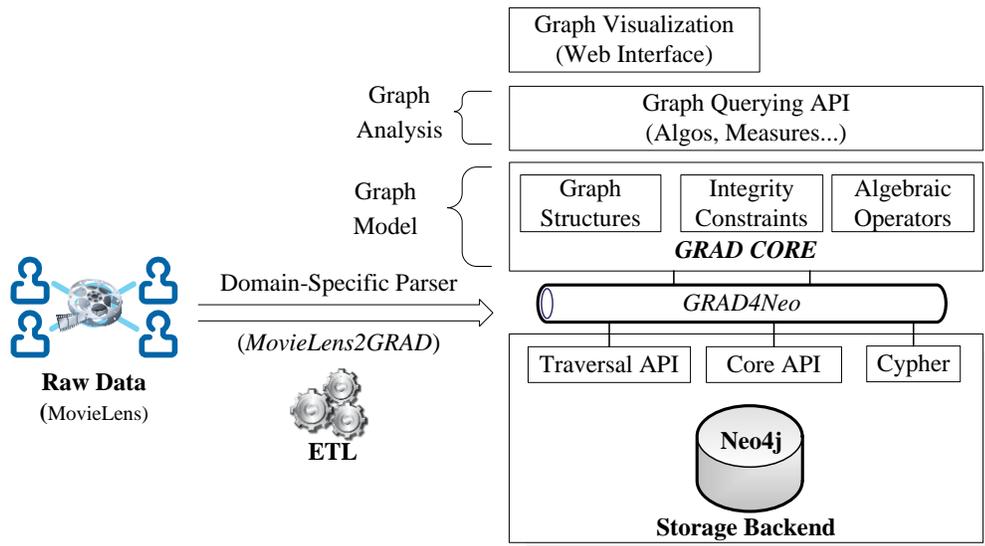

Figure 13. Architecture of a GRAD-based graph analysis system

## 9. RELATED WORK

Graph databases were studied in the nineties within the framework of semi-structured and object-oriented databases (Angles & Gutierrez, 2008). In particular, most research effort was focused on XML and tree structures for semi-structured data management. With the "Big Data" hype and the NoSQL movement, interest in graph databases has regained momentum, partly because many graph-specific issues (such as modeling and efficient querying) have not been efficiently solved by traditional database systems.

As we commented above, two approaches are widely used for modeling and managing graph data. The first one is by leveraging the current models, mainly the relational model. The graph data is represented by two tables, a node table and an edge table. The second one is to build native graph data models and database engines. We discuss next these two approaches.

*Graphs on alternative database models* The advantage of this approach is that once data are loaded in a relational system, it gains the benefits of the well-established relational model, with a smooth integration with a wide range of relational platforms. The relational model was designed to handle structured data such as records and transactions. The data structure and data type are known in advance, and are specified by a fixed schema. However, traditional relational databases fall short of meeting the requirements for complex graph data management. Due to the fundamental difference between the two models, transformation of graph data to the relational model is a manual, complicated process, with a high risk of information loss during the transformation process. This also raises the problem of impedance mismatch at the modeling and querying levels. Supporting changes, extensions or integration of different schemas are difficult tasks. Moreover, the SQL query language cannot target the topology of the graph with queries such as patterns matching, neighborhood or path discovery. The join operation, used to simulate traversals, introduces a heavy workload especially for highly connected tables (Vicknair et al., 2010).

XML data management is closely related to graph data management. Both data representation are based on nodes and edges, and the schema tends to be semi-structured. Techniques for modeling, indexing and querying XML data could be reused as the basis for graph data management. GraphQL for example reuses the FLWOR expressions from XQuery[6]. However, in spite that the XML tree structure is generalized by graphs, which makes modeling more flexible, indexing and querying may become more challenging.

The semantic web stack[7] (comprised of the RDF data model, ontologies and the SPARQL query language) was initially built for enabling interoperability and reasoning on semantic web data. In GRAD, we propose a more natural representation of graph data, with focus on the analysis of structural properties, and without the need for automatic reasoning, inference or ontologies design. Although RDF triple stores are similar to graph databases, their workload and purposes are slightly different. As discussed above, graph databases are optimized for graph traversals, which is not necessarily the case for RDF triple stores, which are rather optimized for handling RDF triples. Another difference is that in graph database properties could be directly added to edges as well as vertices, while this might be more complicated in RDF triple stores and requires some workarounds.

This being said, at the physical level, GRAD could then be implemented using any of the models mentioned above (e.g., XML, RDF, relational). However, the choice at the physical layer will impact the development and modeling effort, and impact the run-time performances.

*Native Graph Databases* In recent years, the trend in developing graph data management systems has shifted to the development of native, relationship-oriented graph databases. Graph databases are optimized for graph traversals. The cost of traversing an edge is constant, and the overall cost of arbitrary navigation through the graph is much lower than the equivalent joins in relational tables (Sadalage & Fowler, 2012). This makes graph databases more suitable for managing complex domains with highly connected data. In addition, the flexibility of native graph models allows the representation of rich structural properties, such as hierarchies and assertions. Finally, implementation aspects such as graph processing, indexing, storage, matching and retrieval which are specifically developed and tuned for graph workloads lead to better performances (Zhao et al., 2007; Tian et al., 2008; He & Singh, 2006).

In industry, a plethora of graph database tools were developed with multiple management features. Neo4j and Spaksee (Martinez-Bazan et al., 2007) are both centralized graph databases. Titan is a distributed graph database. All the three are oriented for online querying of graph data, and support similar features to relational databases such as ACID properties, indexing and query languages. They provide Blueprints-compliant interfaces and are built on top of the property graphs abstraction. HypergraphDB (Iordanov, 2010) is built on hypergraphs, where a single edge might link an arbitrary number of nodes. Trinity (Shao et al., 2013) is a distributed, in-memory key/value store supporting both online and offline graph querying.

---

[6] http://www.w3.org/TR/xquery/
[7] http://www.w3.org/standards/semanticweb/

A comparison of current graph data models was provided by Angles (Angles 2012). These current native graph databases implement different general purpose data models, without a commonly agreed graph modeling approach. Each of these graph models introduces its specific set of data structures and constraints. However, most of the graph databases do not support the three components of a complete database model (Angles, 2012). These databases do not formally define integrity constraints or the graph algebra, and therefore do not provide a formal graph database model to compare with.

In addition to databases, graphs were proposed for extending other components of the data management landscape. For example, Li et al. (2011) and Ghrab et al. (2013) proposed conceptual models for designing and querying graph data warehouse systems. In such settings, GRAD could be reused as the fundamental data model that supports the modeling and querying of the underlying graph database.

*Graph Query Languages* The primary purpose of graph database management systems is to perform online-querying and traversal of large graphs. This is typically useful in social networks analysis, fraud detection and recommendation scenarios. A major key factor in the success of relational databases is the relational algebra and its mapping to the SQL query language. Multiple native graph indexing techniques (He & Singh, 2006; Sun et al., 2012) and query languages (He & Singh, 2010; Dries et al. 2012), were developed to efficiently answer graph-oriented queries (Lee et al., 2012; Wood, 2012). Tian et al. (Tian et al., 2008) proposed a technique K-SNAP for drilling up and down across different aggregation levels on heterogeneous graphs. Zhao et al. studied the issue of querying large heterogeneous information graphs. Their goal is to define an SQL-like declarative language specific for information networks. They designed two operators P-Rank, and SPath for approximate and exact subgraph matching respectively (Zhao et al., 2009).

Querying on current industrial graph databases is mainly done through their API. This approach is more programming-oriented, is application-dependent and requires high programming skills, in contrast with the common declarative approach. Graph databases such as Neo4j defined their own query language, namely Cypher. However, Cypher does not provide an algebraic foundation. It does not support simultaneous querying of collections of graphs. The return-type is tables instead of graphs and therefore do not support piping of graph operations.

To the best of our knowledge, apart from GraphQL, none of the previously cited works proposed a relationally complete algebra. Hence assessing their complexity, expressive power and completeness remains difficult (Angles, 2012). Moreover, as reported by Lee et al., GraphQL is the only algorithm to complete the typical graph queries tested on their comparison (subgraph, clique and path queries), although at a slower performance compared to some of the other tested algorithms (Lee et al., 2012). Another interesting characteristic of GraphQL is that it provides a mapping between the algebraic expressions and FLWOR expressions from XQuery. The query language is declarative, graph-oriented and suited for semi-structured data. A major difference between the original GraphQL and the GRAD Algebra is that GRAD is oriented for the management and querying of a single large graph while GraphQL targets collections of small graphs. Single large graph analysis is usually applied on domains such as social and bibliographic networks with queries centered on pattern matching, reachability and shortest path. Collection of graphs analysis is usually applied on chemical and bioinformatics networks, with analysis oriented for subgraph and supergraph queries (Sakr et al., 2012).

## 10. CONCLUSION

In this paper, we proposed a novel database model for intuitive modeling and effective querying of graph data. GRAD is a complete and native graph database model where relationships are first-class citizens. In GRAD, we proposed advanced graph structures to represent real-world entities. The GRAD algebra is designed such that graphs are the operands and the output of all operators. We defined the integrity constraints to guarantee data consistency.

GRAD is a generic database model that lays the foundations for building specialized graph management systems. It could further be extended to support specific applications requirements. Extensions could be applied at the data structures, algebraic operators, or constraints to fit a certain application requirements. For example, GRAD could be extended to support spatio-temporal graph databases. A further step could be the support of evolving graph databases that take into account both data and schema changes. Graph data warehousing and Master Data Management systems could also benefit from GRAD.

In this paper, we covered the database model, but still the physical layer poses its open challenges. Efficient query processing on large dynamic graphs is an emerging research field that brings many challenges for future research. An optimized graph processing paradigm tuned to GRAD as well as specific storage and indexing mechanisms are still to be developed.